\documentstyle[12pt]{article}
\begin{document}
\begin{center}
\vspace*{1in}
{\Large\bf BACKGROUND FIELDS IN 2+1\\
TOPOLOGICAL GRAVITY}\\[1.5cm]
{\large Andrew Toon\footnote{Supported by the Royal Society England.}\\
University of Oxford\\
Department of Theoretical Physics\\
1 Keble Road\\
Oxford, OX1 3NP\\
England.}
\end{center}
\vspace*{1in}
\begin{abstract}
The partition function of 2+1 Chern-Simons Witten topological gravity has an
attractive physical interpretation in terms of the unbroken and broken phases
of gravity. We make this physical interpretation manifest by using the
background field method.
\end{abstract}
\newpage
\section{Introduction}
It is well known by now that 2+1 dimensional
quantum gravity can be given a gauge theoretic
interpretation in terms of  a Chern-Simons theory with gauge group $ISO(2,1)$,
$SO(3,1)$ or $SO(2,2)$, depending on whether the cosmological constant
$\lambda$ is zero, positive or negative [1,2]. One of the main differences,
however, is the role of non-invertible vierbeins which are forbidden in the
former but are allowed, and even required, in the latter. The Chern-Simons
interpretation of 2+1 quantum gravity, therefore, indicating quite naturally
the existence of an
 unbroken and broken phase of gravity. It turns out that these two phases also
have there place in the physical interpretation
of the partition function of the theory. That is, if a certain operator has no
zero modes then the quantum theory is constrained to the unbroken phase but if
this operator has zero modes then the theory is allowed to escape to the
broken phase where the resulting space-time has a Riemannian interpretation
[2].

The above observations are very interesting but not very manifest. One
discovers them by studying the properties of operators on manifolds with given
topologies. The question to be asked then is
can we us a more physical approach to discover the above observations? One
possibility is the background field method. If the physical interpretation
of the partition function is indeed correct, one would expect the background
fields to mimic correctly this behaviour and thus make manifest this
physical interpretation.

A perhaps naive step in this direction was made in [3]. Here,
Chern-Simons-Witten topological gravity, with non-zero cosmological constant,
was considered on some three manifold $M$. Using the background field method
for the vierbein field $e_{i}^{a}$ only, a one-loop effective action, $\Gamma$,
was obtained of the form:
\begin{equation}
\exp i\Gamma=\det(\lambda
e^{a}_{i})^{-1/2}\det(D_{i}\tilde{D}^{i})\det(D_{i}(\lambda
e^{a}_{j})^{-1}D_{k})^{-1/2}\exp iS[e,\omega,\lambda].
\end{equation}
Here, $e$, $\omega$ and $\lambda$ are the background vierbein, spin connection
and cosmological constant respectively. $D_{i}$ and $\tilde{D}^{j}$ are certain
derivatives required for gauge fixing and $S[e,\omega,\lambda]$ is the initial
action with respect to the background fields with cosmological constant
$\lambda$. It was then argued that the background vierbein $e_{i}^{a}$ must be
everywhere invertible because of the way it appears in the determinants of
equation (1). However, this argument may be a little naive for two reasons.
Firstly, it is clear from equation (1) that the background vierbeins in the
determinants may cancel. If this
is true, it is a bizarre result in that to perform the calculation we have
to assume that $e_{i}^{a}$ is everywhere invertible but at the end of the
calculation everything cancels implying that we have really learned nothing.
The problem, therefore, might be with the particular method used. Secondly, in
obtaining (1), we ignord quantum fluctuations in the spin connection field
which may not be legal.

As an attempt to overcome these difficulties, we will make a more careful
analysis of one-loop effects using the background field method. We will
consider the case of zero cosmological constant as well as the non-zero case.
The case of zero cosmological constant turns out to be quite straight forward
with our results agreeing perfectly with Witten's observations. The case of
non-zero cosmological constant is more difficult
because we want to make sure that any constraints on background fields are not
a consequence of the particular method used.

The outline of this paper is as follows. In section 2, we review the relevant
facts of Chern-Simons-Witten gravity. We then use the background field method
to evaluate an effective action and discuss constraints that appear in this
approach. We then finish with a conclusion.
\section{Gravity in 2+1 dimensions}
In [1,2], Witten constructed a gauge theory of gravity with cosmological
constant $\lambda$ and gauge group $G$ given by:
\begin{equation}
G=\left\{ \begin{array}{ll}
SO(3,1) & \mbox{for $\lambda >0$}\\
ISO(2,1) & \mbox{for $\lambda=0$}\\
SO(2,2) & \mbox{for $\lambda <0.$}
\end{array}
\right.
\end{equation}
The Chern-Simons term on some three manifold $M$ takes the form:
\begin{equation}
S=\frac{1}{2}\int_{M}< A\wedge dA+\frac{2}{3}A\wedge A\wedge A>,
\end{equation}
where $A$ is a Lie algebra valued one form, and $<\; >$ denotes the invariant
quadratic form. Denoting tangent space indices by $i,j,k$ and ^^ ^^ Lorentz"
indices by $a,b,c,$ we expand the gauge field as:
\begin{equation}
A_{i}=e_{i}^{a}P_{a}+\omega_{i}^{a}J_{a},
\end{equation}
where, depending on the sign of $\lambda$, $P_{a}$ and $J_{a}$ are the momentum
and angular momentum generators of the anti-de Sitter, Poincare' or de-Sitter
groups. $e_{i}^{a}$ and $\omega_{i}^{a}$ are the vierbein and spin connection
fields on $M$ respectively. Substituting (4) into (3), we arrive at:
\begin{equation}
S=\frac{1}{2}\int_{M}\epsilon^{ijk}(e_{ia}(\partial_{j}\omega^{a}_{k}-\partial_{k}\omega_{j}^{a})+\epsilon_{abc}e^{a}_{i}\omega^{b}_{j}\omega^{c}_{k}+\frac{1}{3}\lambda\epsilon_{abc}e_{i}^{a}e^{b}_{j}e^{c}_{k}).
\end{equation}
Varying equation (5) with respect to $e$ and $\omega$ gives their respective
equations of motion:
\begin{equation}
\partial_{i}\omega_{j}^{a}-\partial_{j}\omega_{i}^{a}+\epsilon^{abc}(\omega_{ib}\omega_{jc}+\lambda e_{ib}e_{jc})=0,
\end{equation}
\begin{equation}
\partial_{i}e_{j}^{a}-\partial_{j}e^{a}_{i}+\epsilon^{abc}(\omega_{ib}e_{jc}-\omega_{jb}e_{ic})=0.
\end{equation}
Provided that $\det(e^{a}_{i})\neq 0$, the classical equivalence between the
gauge theory and general relativity is seen once one solves $\omega$ in terms
of $e$ via equation (7). When substituted back into equation (5) we find the
Einstein lagrangian with a cosmological constant $\lambda$ [2,4].

Let us now consider the partition function of $S$ on some closed manifold $M$.
Looking at equation (5) we see that its path integral in $e_{i}^{a}$ will
become particularly simple when $\lambda=0$. We therefore discuss this case
since
we expect to be able to calculate the full partition function. Following
Witten [2], we therefore want to evaluate:
\begin{equation}
Z(M)=\int DeD\omega...\exp
i(\frac{1}{2}\int_{M}\epsilon^{ijk}e_{ia}(\partial_{j}\omega^{a}_{k}-\partial_{k}\omega_{j}^{a}+[\omega_{j},\omega_{k}]^{a})+iS_{gf}),
\end{equation}
where $S_{gf}$ is the gauge fixing + ghost term given by:
\begin{equation}
S_{gf}=\int_{M}(uD_{i}e^{i}+vD_{i}\omega^{i}+\bar{f}D_{i}\tilde{D}^{i}f+\bar{g}D_{i}\tilde{D}^{i}g).
\end{equation}
Here $f$ and $g$ are respectively ghosts for the rotations and translations
in $ISO(2,1)$, $\bar{f}$ and $\bar{g}$ are the corresponding antighosts. $u$
and $v$ are the Lagrange multipliers which enforce the gauge conditions:
\begin{equation}
D_{i}\omega^{i}=D_{i}e^{i}=0,
\end{equation}
where $D_{i}=D_{i}^{grav}+[\omega_{(\alpha)i},\;]$, $D_{i}^{grav}$ being the
gravitational covariant derivative with $\omega_{(\alpha)}$ being a flat
connection, and finally $\tilde{D}_{i}=D_{i}^{grav}+[\omega_{i},\;]$ is the
full covariant derivative in the combined gravitational and gauge fields. On
imposing the
gauge conditions (10), one picks a metric $g_{ij}$ on $M$ but this metric is
unrelated to the vierbein which we are regarding as a gauge field. Witten then
distinguishes two cases for the evaluation of $Z(M)$.

In case (1), the topology of the manifold $M$ is such that the moduli space
of flat $SO(2,1)$ connections $\aleph$ consists of finitely many points. In
this case, the partition function is given by:
\begin{equation}
Z(M)=\sum_{\alpha}\frac{(\det \Delta)^{2}}{|\det L|},
\end{equation}
where one is summing over a set of gauge equivalence classes of flat $SO(2,1)$
connections labelled by $\alpha$, $L$ is the bosonic kinetic operator appearing
in equation (8), and $\Delta=D_{i}\tilde{D}^{i}|_{\omega=\omega_{\alpha}}$.

In case (2), the topology of $M$ is such that the moduli space $\aleph$ does
not consist of finitely many points but has positive dimension. In this case
the
operator $L$ has zero modes with the partition function given by:
\begin{equation}
Z(M)=\int_{\Gamma}\frac{(\det \Delta)^{2}}{|\det' L|},
\end{equation}
where $\Gamma$ is the space of classical solutions of the classical field
equations on $M$, and $\det'$ represents the product of the non-zero
eigenvalues.

Let us now consider possible divergences in $Z(M)$ together with their physical
interpretations [2]. If one formulates the quantum theory on some manifold
$M$ that is impossible classically, then the partition function $Z(M)$
corresponds to case (1). Here, the quantum theory never escapes the quantum
regime of
Plankian distances with the partition function being perfectly finite since it
is being dominated by distances of order the Planck scale as there are no zero
modes. That is, case (1) resembles the unbroken phase of gravity.

Case (2) corresponds to formulating the quantum theory on some manifold $M$
with a topology that is allowed classically. In this case, the classical
space-time can be arbitrarly big due to the scale invariance of the classical
equations
of motion.
The partition function $Z(M)$ will then diverge. This divergence being an
infrared divergence which corresponds to the classical region of macroscopic
space-times. This clearly representing the broken phase of the theory.

Our next task is to try and confirm the obove observations using the background
field method.

\section{The background fields}
For $\lambda=0$, the full partition function of the corresponding gauge theory
is precisely the Ray-Singer analytic torsion which is a topological invariant.
This in turn has an attractive physical interpretation in terms of the unbroken
and broken phases of gravity. It is precisly this
physical interpretation we would like to reproduce interms of the background
fields $(e_{i}^{a},\omega_{i}^{a})$.

For the time being we will consider the case $\lambda \neq 0$ and apply the
background field method to evaluate the one-loop contribution to the effective
action of the background fields. The case of zero cosmological constant is
simply
obtained by setting $\lambda=0$. Our basic object of interest is a functional
of background fields which, in principle, contains all the information about
the quantised fields. Our starting point is equation (5). We use the background
field method for the vierbein and spin connection fields to write:
\begin{equation}
e_{i}^{a}\rightarrow e^{a}_{i}+\tilde{e}^{a}_{i},\;
\omega_{i}^{a}\rightarrow\omega_{i}^{a}+\tilde{\omega}_{i}^{a},
\end{equation}
where we are expanding around a non-zero classical solution
$(e^{a}_{i},\omega^{a}_{i})$, and $(\tilde{e}^{a}_{i},\tilde{\omega}_{i}^{a})$
are the quantum fluctuations. At one-loop we require that part of the action,
$S_{2}$, which is at most quadratic in the quantum fields
$(\tilde{e}_{i}^{a},\tilde{\omega}^{a}_{i})$. Thus, making the substitution of
(13) into (5), we obtain:
$$
S_{2}=\frac{1}{2}\int_{M}\epsilon^{ijk}((e_{i}^{a}F_{ajk}(\omega+\tilde{\omega})+\tilde{e}_{i}^{a}F_{ajk}(\omega+\tilde{\omega})+\frac{1}{3}\lambda\epsilon_{abc}e^{a}_{i}e^{b}_{j}e^{c}_{k}$$
\begin{equation}
+\lambda\epsilon_{abc}\tilde{e}^{a}_{i}e^{b}_{j}e^{c}_{k}+\lambda\epsilon_{abc}\tilde{e}^{a}_{i}e^{b}_{j}\tilde{e}^{c}_{k}),
\end{equation}
where $F_{ajk}$ is given by:
\begin{equation}
F_{ajk}(\omega)=\partial_{j}\omega_{ka}-\partial_{k}\omega_{ja}+\epsilon_{abc}\omega^{b}_{j}\omega_{k}^{c}.
\end{equation}
Our next task is to integrate out the quantum fields
$(\tilde{e}_{i}^{a},\tilde{\omega}_{i}^{a})$. Before we can do this, gauge
fixing is required. The one-loop effective action , $\Gamma(e,\omega,\lambda)$,
for the background fields is then given by:
\begin{equation}
\exp i\Gamma=\int D\tilde{e}D\tilde{\omega}...\exp i(S_{2}+S_{gf}),
\end{equation}
where $S_{gf}$ is the gauge + ghost term now given by:
\begin{equation}
S_{gf}=\int_{M}(uD_{i}\tilde{e}^{i}+vD_{i}\tilde{\omega}^{i}+\bar{f}D_{i}\tilde{D}^{i}f+\bar{g}D_{i}\tilde{D}^{i}g).
\end{equation}
As in equation (9),  $f$ and $g$ are respectvely ghosts for the $\tilde{e}$ and
$\tilde{\omega}$ fields, $\bar{f}$ and $\bar{g}$ are the corresponding
antighosts. $u$ and $v$ are Lagrange multipliers which enforce the gauge
conditions:
\begin{equation}
D_{i}\tilde{e}^{i}=D_{i}\tilde{\omega}^{i}=0,
\end{equation}
where now, $D_{i}=D_{i}^{grav}+[\omega_{i},\;]$, $D_{i}^{grav}$ being the
gravitational covariant derivative, $\omega_{i}$ being the background field
connection, which we will be assuming obeys the background field equations, and
now $\tilde{D}_{i}=D_{i}$.

What we would like to do is expand $\tilde{e}$ interms of eigenfunctions of
some operator but since the background field $e$ is a classical object, we
cannot regard it as a operator and so doing this will give us little
information
regarding the background field $e_{i}^{a}$. We will therefore proceed by using
formal
manipulations of the path integrals.

Let us first practice with the case $\lambda=0$. Here, the one-loop effective
action becomes the full effective action since we have only linear terms in
$e$. Performing the $\tilde{e}$, $v$, $f$ and $g$ integrals, gives:
$$
\exp i\Gamma(e,\omega,\lambda=0)=\int
D\tilde{\omega}Du\delta(\frac{1}{2}\epsilon^{ijk}F_{ajk}(\omega+\tilde{\omega})+D^{i}u_{a})$$
\begin{equation}
\times\delta(D^{i}\tilde{\omega}_{i}^{a})\det(\Delta).\exp
i\frac{1}{2}\int_{M}\epsilon^{ijk}e^{a}_{i}F_{ajk}(\omega+\tilde{\omega}).
\end{equation}
In order to evaluate the $\tilde{\omega}$ integral, we borrow the method of
reference [2]. We start by demanding that the background field $\omega_{i}^{a}$
obeys the classical equation of motion given by equation (6). This means that:
\begin{equation}
F_{aij}(\omega)=0.
\end{equation}
In order that $F_{aij}(\omega+\tilde{\omega})=0$, which follows from equation
(19), we find to lowest order in $\tilde{\omega}$ that:
\begin{equation}
\frac{1}{2}\epsilon^{ijk}F_{ajk}(\omega+\tilde{\omega})=\epsilon^{ijk}
D_{(\omega)j}\tilde{\omega}_{ak},
\end{equation}
where $D_{(\omega)j}$ is the covariant derivative with respect to the
background flat connection $\omega_{i}^{a}$. Assuming that $D_{(\omega)j}$ has
no zero modes, we can
change variables in the path integral from $\tilde{\omega}$ to
$\epsilon^{ijk}D_{(\omega)i}\tilde{\omega}_{jk}$ to obtain:
\begin{equation}
\exp i\Gamma=\int Du\frac{(\det\Delta)^{2}}{|\det D_{(\omega)j}|}\exp
i\int_{M}e_{i}^{a}D^{i}u_{a}.
\end{equation}
Finally, performing the $u$ integral gives, assuming that $D^{i}$ has no zero
modes:
\begin{equation}
\exp i\Gamma=\frac{(\det\Delta)^{2}}{|\det D_{(\omega)j}|.|\det
D^{i}|}\delta(e_{i}^{a}).
\end{equation}
We see that the background field $e_{i}^{a}$ is manifestly constrained to be
zero. That is, when the operators $D_{(\omega)j}$ and $D^{i}$ have no zero
modes, the theory is constrained to the unbroken phase\footnote{Note that the
operator $L$ in equation (11) is
given by $L=\left(\begin{array}{ll}
         \epsilon^{ijk}D_{(\alpha)j} & D^{i}\\
         -D^{k} & 0
\end{array}
\right)$}.

Lets now consider the opposite situation in which the operators $D_{(\omega)}$
and $D^{i}$ have zero
modes. In this case it is clear what we have to do. Starting from equation
(19), we will, as in reference [2], not integrate over these zero modes which
we label by $\tilde{e}_{0}$. Thus equation (19) becomes:
\begin{equation}
\exp i\Gamma=\int D\tilde{e}_{0}Du\frac{(\det\Delta)^{2}}{|\det'
D_{(\omega)j}|}\exp i\int_{M}e_{i}^{a}D^{i}u_{a},
\end{equation}
where $\det'$ represents the product of non-zero eigenvalues. Performing the
$u$ integral, noting that we cannot use the substitution $u_{a}\rightarrow
D^{i}u_{a}$ in the path integral which was used in the case when $D$ has no
zero modes, gives:
\begin{equation}
\exp i\Gamma=\int D\tilde{e}_{0}\frac{(\det \Delta)^{2}}{|\det
'D_{(\omega)j}|}\delta(D^{i}e_{i}^{a}).
\end{equation}
Clearly, the background field $e_{i}^{a}$ is now constrained to:
\begin{equation}
D^{i}e_{i}^{a}=0,
\end{equation}
that is, equation (26) has non-zero solutions. What does this equation mean?
When gravity is discussed in the first order formalism, it is always possible
to define the affine connection by postulating that the vierbein is covariantly
constant [5], that is:
\begin{equation}
D_{i}e_{j}^{a}=\partial_{i}e_{j}^{a}-\Gamma^{k}_{ji}e^{a}_{k}+\epsilon^{abc}\omega_{ib}e_{jc}=0.
\end{equation}
{}From this equation we can deduce the standard metric connection of general
relativity. Equation (26) is saying much the same thing when we recognise that
(26)
becomes:
\begin{equation}
g^{ij}D_{i}e_{j}^{a}=0,
\end{equation}
where $g_{ij}$ is the gauge fixing metric which is also consistant with
equation (27). What all this is saying is that the non-zero background field
$e_{i}^{a}$ is a solution to Einstein's equations and thus corresponds to the
broken phase of gravity. This of course confirming the observations of equation
(12).

Let us now consider the case of $\lambda\neq 0$. Looking back at equation (14),
we see that we now have a quadratic and a linear term in the quantum field
$\tilde{e}^{a}_{i}$. Integrating over this field will straight away imply
non-trivial constraints on the background field $e^{a}_{i}$ since we will have
factors like
$(\det e_{i}^{a})^{-1/2}$ and $(e_{i}^{a})^{-1}$, which would imply that the
background fields must be everywhere invertible on $M$ for the one-loop
partition
function to make sense. Of course, this may not be true. It is quite possible
that the determinants may cancel as discussed in the introduction. Thus by
demanding that the background field be everywhere invertible might just be
a facit of the particular method used. For this reason we will follow a
different route in evaluating (19). We will do this by assuming that the
partition function (19) is still dominated by flat connections as for the case
$\lambda=0$.
This certainly seems reasonable for very small $\lambda$ which in any case must
be true for the one-loop partition function to be a good approximation
to the full theory.  Starting from equation (19), we will firstly
perform the $\tilde{\omega}$ integral together with the $v$, $f$ and $g$
integrals. Assuming then that the one-loop
partition function is dominated by flat connections, we can use equation (21).
Assuming that $D_{(\omega)}$ has no zero modes leads to:
$$
\exp i\Gamma=\int D\tilde{e}Du\frac{(\det\Delta)^{2}}{|\det
D_{(\omega)}|}\delta(e_{i}^{a}+
\tilde{e}_{i}^{a})$$
\begin{equation}
\times\exp
i\int_{M}\epsilon^{ijk}(\frac{1}{3}\lambda\epsilon_{abc}e_{i}^{a}e_{j}^{b}e_{k}^{c}+\lambda\epsilon_{abc}\tilde{e}^{a}_{i}e_{j}^{b}e_{k}^{c}+\lambda\epsilon_{abc}\tilde{e}_{i}^{a}e_{j}^{b}\tilde{e}_{k}^{c})+u_{a}D^{i}\tilde{e}_{i}^{a}.
\end{equation}
It is now quite clear that we may perform the $\tilde{e}_{i}^{a}$ integral, and
also performing the $u$ integral as in the $\lambda=0$ case, gives:
\begin{equation}
\exp i\Gamma=\frac{(\det \Delta)^{2}}{|\det D_{(\omega)j}|.|\det
D^{i}|}\delta(e_{i}^{a})
\exp i\int_{M}\epsilon^{ijk}\epsilon_{abc}\frac{1}{3}\lambda
e_{i}^{a}e_{j}^{b}e_{k}^{c}.
\end{equation}
Again, we see that the background field is constrained to the unbroken phase as
with the $\lambda=0$ case.

Consider now the case when $D_{(\omega)}^{i}=D^{i}$ has zero modes. Since we
are now performing the
$\tilde{\omega}_{i}^{a}$ integral first, we cannot follow the corresponding
situation  as in $\lambda=0$ since there we considered the operator
$D_{(\omega)j}$ in terms
of eigenfunctions of $\tilde{e}_{i}^{a}$. So it would not make sense in this
case. We therefore expand $\tilde{\omega}_{i}^{a}$ in terms of eigenfunctions
of
$D$, leaving un-integrated these zero modes $\tilde{\omega}_{0}$, we find on
performing the $u$ integral:
\begin{equation}
\exp i\Gamma=\int D\tilde{\omega}_{0}\frac{(\det \Delta)^{2}}{|\det
'D|}\delta(D^{i}e_{i}^{a})\exp
i\int_{M}\frac{1}{3}\lambda\epsilon^{ijk}\epsilon_{abc}e_{i}^{a}e_{j}^{b}e_{k}^{c},
\end{equation}
where the same conclusions follow as for the $\lambda=0$ case.
\section{Conclusion}
The point of this paper as been to try and reproduce Witten's physical
interpretation of the partition function of 2+1 topological gravity in terms of
background
fields. The case $\lambda=0$ was quite straight forward. Here, the background
field $e_{i}^{a}$ was shown to mimic correctly the unbroken and broken phases
of gravity.
The case $\lambda\neq 0$ was a little more suttle. If one attempted to
integrate over the quantum field $\tilde{e}_{i}^{a}$ first, then one must
assume that
the background field $e_{i}^{a}$ is everywhere invertible on $M$ to make sense
of the expression. As discussed in the introduction, this may be a little naive
and at the end of the calculation, all constraints may cancel implying that
we really cannot make sense of this procedure. To over come these difficulties,
we chose in this case to integrate over $\tilde{\omega}_{i}^{a}$ first where
now we must think of expanding the quantum field $\tilde{\omega}_{i}^{a}$ in
terms of eigenfunctions of $D$. We then assumed that the one-loop partition
function is dominated by flat connections as with the $\lambda=0$ case. We
note that this is different from assuming that the one-loop partition function
is dominated by classical solutions given by equation (6). The above assumption
certainly makes sense if $\lambda$ is very small, which as to be true for the
one-loop partition function to be a good representation of the full theory.
We then followed a similair procedure as for the $\lambda=0$ case and obtained
similar results. This of course should be no surprise since we expect to
recover the $\lambda=0$ results in the limit $\lambda\rightarrow 0$.

Along some what different lines, 2+1 topological gravity was investigated by
introducing a space-time ^^ ^^ metric" $G_{ij}$ as a collective coordinate [6].
It was shown that $G_{ij}$ also mimiced correctly the two phases of quantum
gravity giving encouraging evidence that $G_{ij}$ is indeed some space-time
metric. Thus, wheather working with background vierbein and spin connection
fields or a space-time ^^ ^^ metric" $G_{ij}$, we find that we are able to
make the broken and unbroken phases manifest. It would be interesting to
understand the relationship between these two approaches in more detail.

I would like to thank Gareth Williams for discussions. This work was supported
by the Royal Society England.
\vspace*{1cm}
\begin{center}
{\bf References}
\end{center}

[1] E. Witten, Nucl. Phys. B311 (1988/89) 46.\\

[2] E. Witten, Nucl. Phys. B323 (1989) 113.\\

[3] A. Toon, Phys. Rev. D47 (1993) 2435.\\

[4] A. Chamseddine, Nucl. Phys. B346 (1990) 213.\\

[5] E. Alvarez, Rev. Mod. Phys. Vol. 61, No. 3 (1989) 561.\\

[6] G. Williams, A. Toon, University of Shizuoka preprint, SU-LA-92-06.
\end{document}